# Zero phase delay induced by wavefront modulation in photonic crystals

[1]Dong Guoyan, [1]Zhou Ji* and [2]Cai Luzhong

[1]State Key Lab of New Ceramics and Fine Processing, Department of Materials Science and Engineering, Tsinghua University, Beijing 100084, P. R. China

[2]Department of Optics, Shandong University, Jinan, 250100, P. R. China

*zhouji@tsinghua.edu.cn

A new mechanism for generation of efficient zero phase delay of electromagnetic wave propagation based on wavefront modulation is investigated in this paper. Both numerical simulations and experiment have demonstrated the zero phase delay behaviors of wave propagation in a two-dimensional triangular photonic crystal. The zero phase delay propagation, independent of the propagating distance, is attributed to an invariable wavefront modulation of photonic crystal in the direction of wave propagation, where the phase velocity is perpendicular to the group velocity with parallel wavefronts (or phasefronts) extending along the direction of energy flow. This effect can be extended to the three-dimensional cases or other artificially engineered materials, and may open a new route to obtain perfect zero-phase-delay propagation for electromagnetic wave instead of using zero-index or zero-averaged-index materials and have significant potential in many applications.

Recently, great efforts have been made to construct materials with zero or near-zero-$n$ with quasi-uniform phase and infinite wavelength [1-16]. Zero-$n$ materials have a series of exciting potential applications, such as wavefront reshaping [10], beam self-collimation [5, 13], extremely convergent lenses [17], etc. One of their most important applications is the optical links in lumped nanophotonic circuits that can guide light over hundreds of wavelengths without introducing phase variations so as to reduce the unwanted effects of frequency dispersion. Several different strategies have been applied to realize the zero-$n$ material (or zero permittivity $\varepsilon$). One of them was to use metallic metamaterial structures with effective permittivity and/or permeability near zero [7, 8]. Usually these materials suffer from strong resonance loss and hence the greatly deteriorated transmission efficiency. Alternative approaches include the microwave waveguides below cutoff [9], the combination of negative- and positive-index materials[13] or the periodic superlattice formed by alternating strips of positive index homogeneous dielectric media and negative index photonic crystals (PhCs) [18] with zero phase accumulation of a wave travelling through the whole superlattice. However, all these configurations require high fabrication precision and complicated architecture. Therefore, it is preferred to find an efficient and simpler way to achieve zero phase delay propagation of electromagnetic wave (EMW).

Wavefront modulation supplies a novel method to realize equal phase transmission of EMW other than conventional approach by using materials with zero-index or zero-average-index. In physics, a wavefront is the locus of points having the same phase and EMW's propagation in media always undergoes phase shift with the wavefront perpendicular to the direction of energy flow. A plane wave traveling in arbitrary direction can be described as

$$E(\mathbf{r}) = A\cos(\mathbf{k} \cdot \mathbf{r} - \omega t + \varphi_0), \tag{1}$$

with the complex amplitude denoted by

$$\widetilde{E}(\mathbf{r}) = A\exp[i(\mathbf{k} \cdot \mathbf{r} + \varphi_0)], \tag{2}$$

here $A$ is wave amplitude, $\mathbf{k}$ is wave vector pointing to the direction of phase velocity $v_{ph}$ with the magnitude of $|\mathbf{k}|=2\pi/\lambda$, $\lambda$ is the wavelength, and $\mathbf{r}$ is position vector, $\varphi_0$ is initial phase. In spatial domain, the phase shift between two different positions is determined by the spatial phase factor $\mathbf{k} \cdot \mathbf{r}$. If the origin of coordinate system is defined on the propagation path, the position vector $\mathbf{r}$ points to the direction of energy flow (i.e. the direction of group velocity $v_{gr}$). When EMW travelling in a conventional material with an ordinary refractive index, i.e. the right-handed (RH) material, the wavefronts travel away from the source with $\mathbf{k} \cdot v_{gr} > 0$ and gain a positive relative phase. In a left-handed (LH) material, however, the wavefronts travel toward the source with $\mathbf{k} \cdot v_{gr} < 0$ and accumulate a negative relative phase. Regardless of the propagation mode, the energy flow essentially remains away from the source along the transmission direction. Supposing the condition of $\mathbf{k} \cdot v_{gr}=0$ is satisfied, the wave vector $\mathbf{k}$ pointing to the normal of wavefront is perpendicular to the energy flow with the planar phase fronts extending along the transmission direction, which means that the spatial phase is constant along the energy flow, in other words, zero phase delay of EMW propagation can be realized with the method of wavefront modulation to satisfy the condition of $\mathbf{k} \cdot \mathbf{r} =0$. On the contrary, in conventional EMW propagation, it is difficult to dramatically modulate wavefronts parallel to the energy flow in homogeneous materials.

Photonic crystals [19, 20] are artificial materials with spatially periodical dielectric functions, which can be used to engineer the optical and electromagnetic properties through the geometry of their unit cells. The tailored diffraction gives rise to distinct optical phenomena, such as high-reflecting omni-directional mirrors [21], low-loss-waveguiding [22], super lens [23] and effective negative refraction [24, 25]. In contrast to strong resonance mechanism in metamaterials, the hypernormal optical phenomena of PhC are based on the special dispersion relations of photonic bands with the smaller loss than the former. The method of wavevector diagram is generally used to study the properties of beam propagation in PhCs, in which the k-conservation rule is generalized to satisfy the boundary condition at the interface of PhC with $k_{\parallel}=\omega/c\sin\theta_{inc}$, and the propagation direction of refracted wave is oriented to the group velocity vector $v_{gr} = \nabla_k\omega$ which is normal to the equal frequency contour (EFC). Due to the large design flexibility of PhC, the EFCs of PhC dominated by the photonic band structure usually present some peculiar shape at certain frequencies to make the propagation direction of refracted wave not parallel to the wave vector, and then induce various transmission phenomena with special optical properties. Hence, PhC can be one optimum candidate for wavefront modulation to achieve zero phase delay of EMW propagation.

As an example, a two-dimensional (2D) triangular lattice PhC is proposed in this work to study this novel effect. This PhC slab is composed of low-lossy dielectric rods with $\varepsilon_r$=10 closely packed in air with a center-to-center separation of $a$ ($a$ is the lattice constant), as shown in Fig. 1(a). By utilizing the plane-wave expansion method, we calculate the lowest four bands for TM polarization with electric field $\boldsymbol{E}$ parallel to the dielectric rods as shown in Fig. 1(b) with the normalized frequency of $a/\lambda$, where the inset represents the first Brillouin zone of this triangular PhC with three high-symmetry lattice points, and the point Γ is the Brillouin zone center. Different from the other bands, the fourth band denoted by the red solid line has the more undulations which imply the more complicated optical properties.

In order to clearly analyze the wave optics characteristics of this PhC, the EFC plot of the fourth band is calculated and shown in the right part of Fig. 2 with the highlighted black bold EFCs at the frequency of 0.374 $a/\lambda$, which look like six leaves gathering around the center point Γ. Supposing the source plane wave is incident from air with the incident angle of $\theta_{inc}$ =30° upon the interface between air and the PhC slab with the surfaces along the ΓK direction, the corresponding wave vector diagram is illustrated Fig. 2, where the blue circle represents the air EFC of $\omega$ = 0.374 $a/\lambda$, the black arrow denotes the incident wave vector $\boldsymbol{k}_i$ in air, the orange arrows stand for refractive wave vectors $\boldsymbol{k}_r$ which are always perpendicular to the phase front, the dashed line means the conservation of the parallel components of wave vectors which validly intersects the EFC of $\omega$ = 0.374 $a/\lambda$ twice at different points A and B, and the red and green arrows demonstrate the group velocity directions of two refracted waves A and B, respectively. By the definition of group velocity $\boldsymbol{v}_{gr} = \nabla_k \omega$, the group velocity vector is oriented perpendicular to the EFC surface in the frequency-increasing direction. By rigorous theoretical analysis, we found that the group velocity of refracted wave A is perpendicular to its wave vector with $\boldsymbol{k} \cdot \boldsymbol{v}_{gr} = 0$, which indicates that the condition of zero phase shift can be satisfied in this case. With the frequency increasing, the surrounding EFCs will intersect at the center point. Numerous calculations show that the condition of $\boldsymbol{k} \cdot \boldsymbol{v}_{gr} = 0$ can also be satisfied in other cases, such as the incident EMW of 0.376 $a/\lambda$ with $\theta_{inc}$ =10°.

Numerical simulations have been introduced to demonstrate the predicted analytical results. A continuous Gaussian wave source which can be regarded as a plane wave with the spatial width of 6.5$a$ is located in front of the PhC slab with the thickness of 15 layers. As shown in Fig. 3(a), when the source wave with $\omega$ = 0.374$a/\lambda$ is incident upon the ΓK surface of this PhC slab with $\theta_{inc}$ =30°, the wavefront of refracted beam A is modulated by the periodic structure to recombine in the PhC slab with the parallel wavefronts extending along the propagation direction, whose spatial period is approximately 5.2$a$ in the normal direction. Fig. 3(b) shows the simulation of electric field distribution in the PhC slab with the incident EMW of 0.376$a/\lambda$ at $\theta_{inc}$=10°, whose modulated wavefronts in the PhC are also parallel to the transmission direction with a larger spatial period of about 14$a$. For general optical systems, phasefront and wavefront are identical as long as the nominal wavefront deviation remains relatively small.

To further verify the above mentioned peculiar phenomena experimentally, we design a sample of triangular lattice PhC slab formed by high pure $Al_2O_3$ ceramic rods with dielectric constant $\varepsilon_r$=10, relative permeability 1 and loss tangent 0 in this study. The geometric parameters of these $Al_2O_3$ ceramic rods are designed as diameter $d$=10mm, height $h$=10mm and lattice constant $a$=10mm, with the same lattice structure as in the simulations above. The sample is placed in a near-field scanning system (microwave planar waveguide) [26], as shown in Fig. 4, where the upper and lower metal plates form the planar waveguide to ensure transverse electromagnetic (TEM) mode invariable between the plates along the z-axis. A rectangular waveguide with the cross-sectional dimensions of 22.86mm ×10.16mm is utilized in the lower plate as the X-band waveguide adapter to excite a plane wave with the electric field ***E*** perpendicular to the upper and lower conducting plates. In the x-band region (8-12GHz), only the dominant mode of $TE_{10}$ can propagate in the waveguide. A detecting probe is mounted in the upper plate to measure the amplitude and phase in the local electric field. The feeding and detecting probes are connected to the output and input ports of the Vector Network Analyzer (VNA) (Agilent ENA5071C). The lower metal plate is carried by the 2D moving stage which can be controlled by a computer to move in x and y directions with a scanning step of 4 mm, so that we can measure EM field distributions within an area.

Next, two experiments are conducted to prove the previously mentioned effects of zero phase delay in PhC by modulating the wavefront of transmitted EMW. We show in Fig. 5 the measured electric fields (real part and phase) for each of the cases examined. In the first experiment, the source wave at 11.213GHz (i.e. relative frequency of 0.374$a/\lambda$) is incident upon the PhC slab with the incident angle of $\theta_{inc}$=30°, both the amplitude and phase internal fields of this PhC sample are measured, allowing us to form an image that can be compared directly with the numerical simulations. In Fig. 5(a), the refracted EMW propagating in the PhC exhibits desired perfect planar wavefronts at the entrance position with the spatial periodic ≈5.2$a$, with the further propagation of light in the sample, some wavefront deviations arise from the constructive interference and diffraction of the finite sample area. Since the phase at a given detection location is referenced to the output port of the VNA, it is necessary to scan the phase contrast field in the sample to match up with the analytical simulations. As shown in Fig. 5(b), the negative phase described with the blue area (2) in the PhC slab connects the incident wave and the transmitted wave with the same phase, and the positive phase front is denoted by the red area (1), with the phase contrast mode sliding perpendicular to the energy flow as the black arrow shows. In contrast to the wavefront aberration in Fig. 5(a), the phase fronts in Fig. 5(b) are seen to be considerably more coherent with the simulation results. Due to the nearly invariable phase along the transmission direction, the phase difference between the incident wave and transmitted wave is almost equal to zero just like one continue wave without spatial separation. Further experimental study find that the amplitude and phase distribution of the internal EM field present a dramatic dependence on the frequency and the incident angle. When the source wave of 11.368 GHz is incident at a small incident angle of 10°, the measured electric fields (real part and phase) are demonstrated in Fig. 5(c, d) with a larger spatial period of about 14.1$a$. The negative and positive phasefronts can be more clearly distinguished by the blue area (1) and red area (2) in phase contrast image of Fig. 5(d). In these cases, the combined effects of wavefront (or phasefront) modulation in the PhC sample agree well with the simulation results, which demonstrate that the method of wavefront modulation is an efficient method to realize zero phase delay of EMW propagation.

Compared with the uniform phase distribution in zero-index materials, parallel phasefronts extend along the transmission direction of EMW in the PhC slab, in the spatial domain the phase difference between arbitrary spatial locations in the propagation direction is static and equal to zero, while remaining dynamic in the time domain, thus allowing energy transport. In contrast to the longitudinal direction of the path-averaged zero-index nanofabricated superlattices in Ref.[18], the direction of EMW propagating with zero phase delay in the PhC slab can be easily manipulated by adjusting the incident angle, and the effect of zero phase delay can be realized within a wide scope of incident angle from 28° to 40° in this case. Different from the finite plasma frequency of such materials follow Drude or Drude-Lorenz dispersion models[10, 27, 28], the zero phase delay propagation of EMW based on the wavefront modulation can be realized at different frequencies for a certain PhC as long as the ***k*** · ***v***$_{gr}$ = 0 condition is satisfied. These unique features of this effect may show great impacts on both fundamental physics and optical device applications.

In conclusion, we have proposed a new mechanism to realize the zero phase delay for EMW propagation in media based on a novel feature of wavefront modulation of photonic crystals. A triangular lattice PhC can make the wavefront (or phasefront) of EMW parallel to the transmission direction and achieve the zero phase shift between the incident and transmitted EMWs. This effect can also be further extended to the 3D case by modulating the wavefront of EMW to satisfy the condition of ***k*** · ***v***$_{gr}$ = 0. Since the PhC structure can be engineered easily and have large design flexibility, engineered control of phase delay in PhCs may be implemented in chip-scale transmission lines, interferometers with deterministic phase array and dispersion control, and therefore has significant technological potential in phase-insensitive image processing, lumped elements in optoelectronics, information processing.

We gratefully acknowledge the financial support from National Natural Science Foundation (51102148, 51032003 and 11274198), National High Technology Research and Development Program of China (863 Program) (2012AA030703), China Post-doctoral Research Foundation (20110490345, 2012T50087) and Shandong Natural Science Foundation (ZR2010AM025, ZR2011FQ011).

# Figure Caption

FIG. 1. (Color online) (a) Schematic of the 2D triangular PhC with $Al_2O_3$ ceramic rods closely packed in air; (b) Band diagram of this PhC for TM polarization.

FIG. 2. (Color online) EFCs plot of the fourth band with the wave vector diagram at the frequency of 0.374a/$\lambda$ with $\theta_{inc}$ =30°.

FIG. 3. (Color online) FDTD simulations of electric field distribution in the PhC slab with the incident beams of (a) $\omega$=0.374 $a/\lambda$ with $\theta_{inc}$ =30° and (b) $\omega$=0.376$a/\lambda$ with $\theta_{inc}$ =10°.

FIG. 4. (Color online) Photograph of the measurement system.

FIG. 5. (Color online) The measured spatial mappings of (a) electric field and (b) phase contrast image at 11.213GHz with $\theta_{inc}$=30°; (c)the electric field and (b) phase contrast image at 11.368GHz with $\theta_{inc}$=10°.

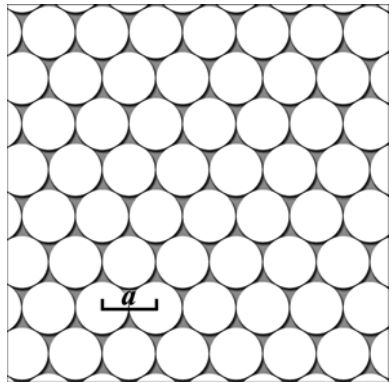

(a)

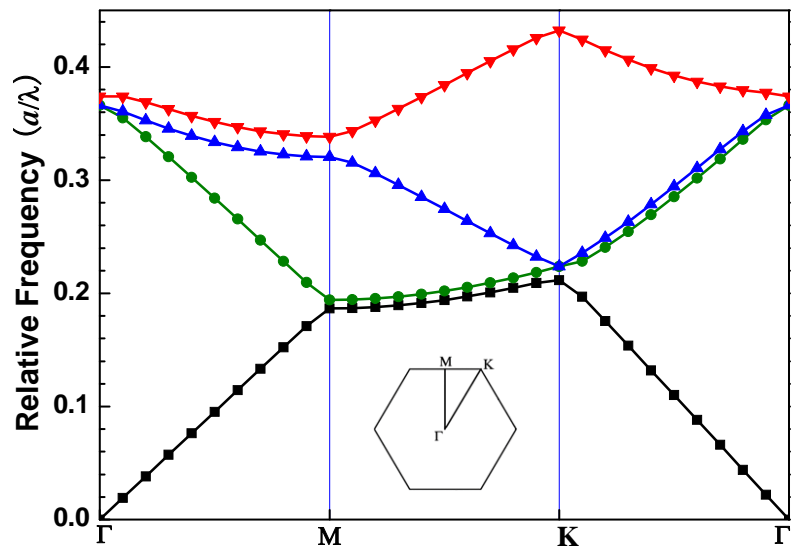

(b)

FIG. 1

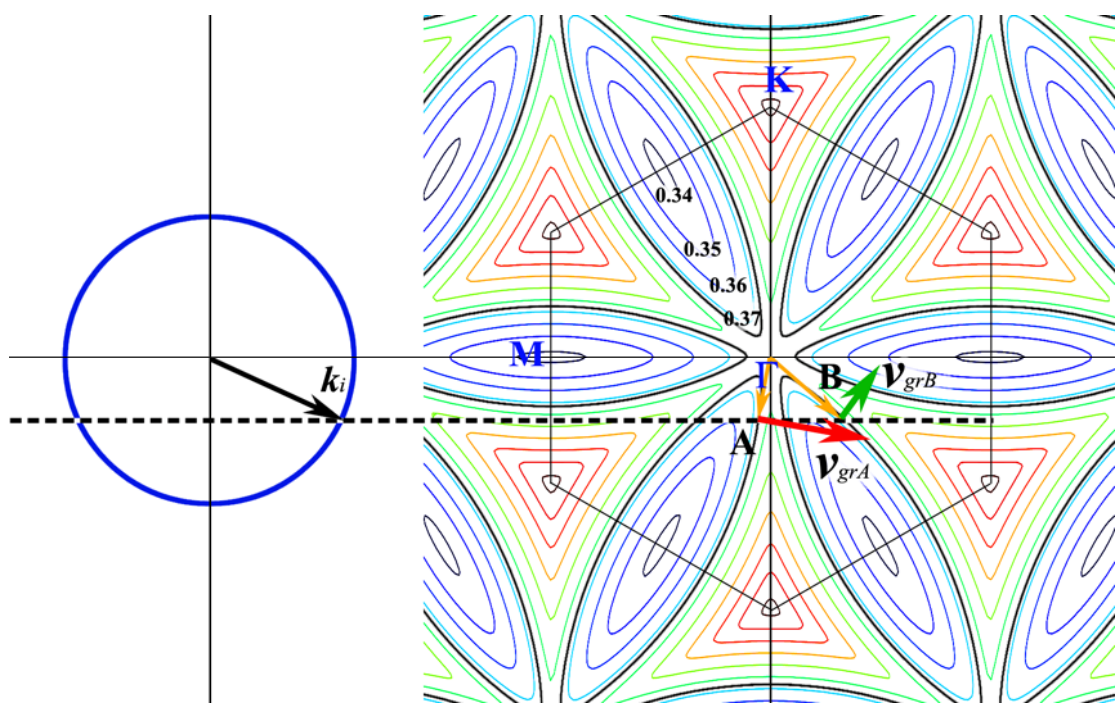

FIG. 2

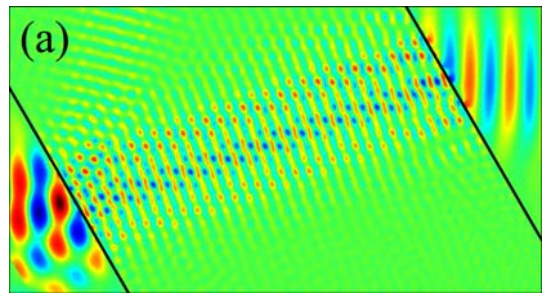

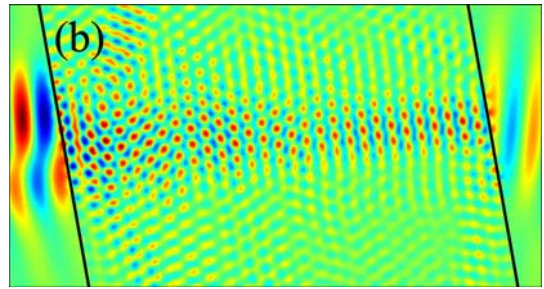

FIG. 3

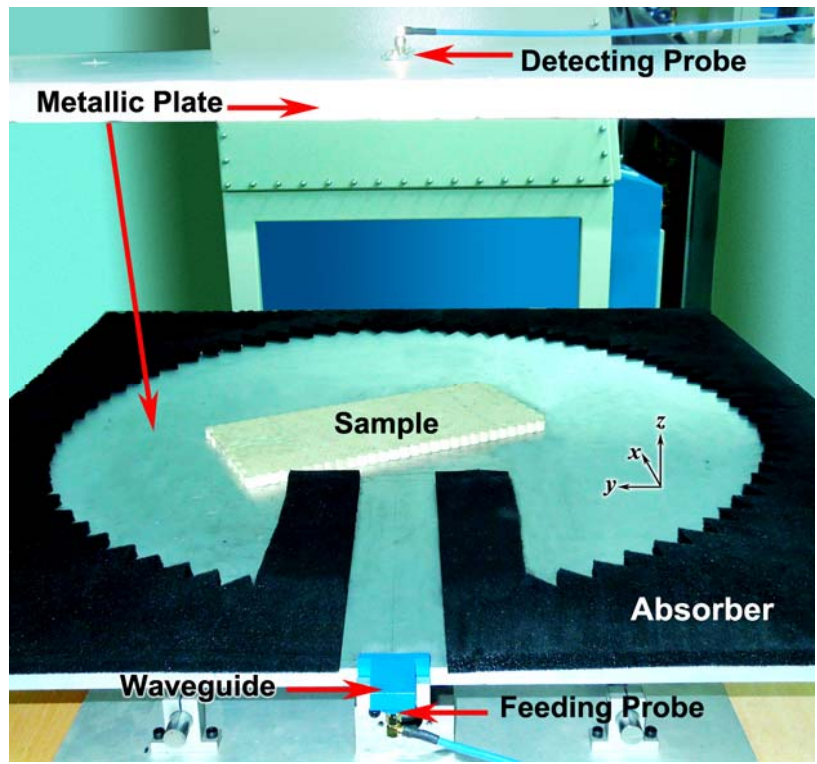

FIG. 4

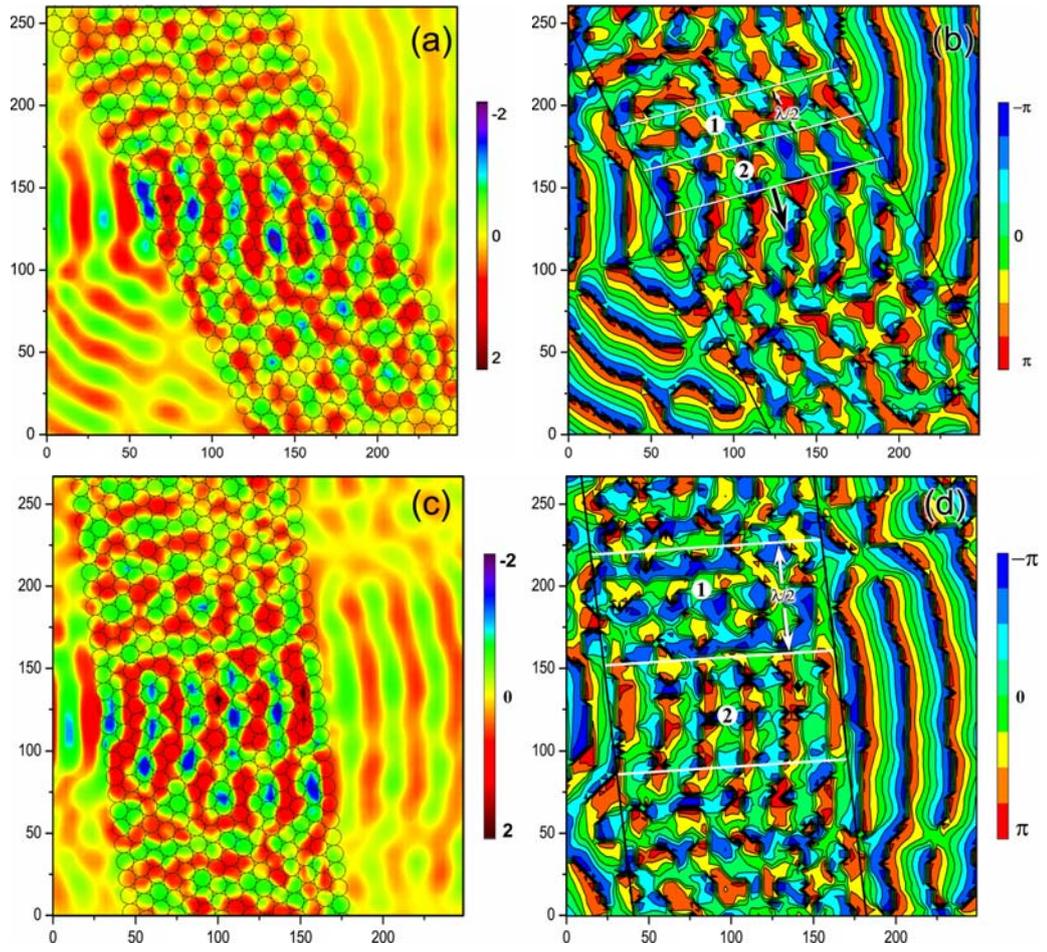

FIG. 5